\documentstyle[12pt,epsf]{article}

\begin{document}

\renewcommand{\theequation}{\thesection.\arabic{equation}}
\newcommand{\reseteqnum}{\setcounter{equation}{0}}

\newcommand{\C}{{\bf C}}
\newcommand{\Z}{{\bf Z}}
\newcommand{\ZnZn}{{\bf Z}_n \times {\bf Z}_n}
\newcommand{\ZtZt}{{\bf Z}_3 \times {\bf Z}_3}

\def\f#1#2{\textstyle{#1\over #2}}
\def\drawbox#1#2{\hrule height#2pt
        \hbox{\vrule width#2pt height#1pt \kern#1pt
              \vrule width#2pt}
              \hrule height#2pt}
\def\Fund#1#2{\vcenter{\vbox{\drawbox{#1}{#2}}}}
\def\Asym#1#2{\vcenter{\vbox{\drawbox{#1}{#2}
              \kern-#2pt       
              \drawbox{#1}{#2}}}}
\def\fund{\Fund{6.5}{0.4}}

\title{
\hfill
\parbox{4cm}{\normalsize UT-KOMABA 99-21\\hep-th/9912273}\\
\vspace{2cm}
Brane Cube Realization of\\
Three-dimensional Nonabelian Orbifolds
\vspace{1.5cm}}
\author{Tomomi Muto\thanks{e-mail address:
\tt muto@hep1.c.u-tokyo.ac.jp}
\vspace{0.5cm}\\
{\normalsize\em Institute of Physics}\\
{\normalsize\em University of Tokyo, Komaba, Tokyo 153, Japan}}
\date{\normalsize}
\maketitle
\vspace{1cm}

\begin{abstract}
\normalsize

We study D-branes
on three-dimensional orbifolds ${\bf C}^3/\Gamma$
where $\Gamma$ are finite subgroups of $SU(3)$.
The quiver diagram of $\ZnZn \in SU(3)$ can be expressed
in three-dimensional form.
According to the correspondence between quiver diagrams
and brane configurations,
we construct a brane configuration for ${\bf C}^3/\ZnZn$
which has essentially three-dimensional structrue.
Brane configurations for nonabelian orbifolds
$\C^3/\Delta(3n^2)$ and $\C^3/\Delta(6n^2)$
are obtained from that for $\C^3/\ZnZn$
by certain quotienting procedure.
\end{abstract}

\newpage
\section{Introduction}

During recent years supersymmetric field theories
have been investigated using D-branes on a singular manifold
such as orbifolds and a conifold.
In this setup field theories arise as world volume theories
of the D-branes.
Aspects of field theories in question are encoded in
geometric information of the singularity.

On the other hand there has been another approach
of construction of supersymmetric field theories
using branes which appear in string theory.
In this framework quantities of field theories
are determined by configurations of branes.
This approach has an advantage that aspects of
field theories can be visualized.
For some cases relations between the two approaches
have been discussed,
and investigations on brane configurations are also helpful
to the study of geometries around singularities probed by D-branes.

In this paper we study
brane realizations of supersymmetric field theories
corresponding to D-branes on an orbifold ${\bf C}^3/\Gamma$
with $\Gamma$ a nonabelian finite subgroup of $SU(3)$.
Finite subgroups of $SU(3)$ are classified into
ADE like series just like the finite subgroups of $SU(2)$.
For A-type subgroups, which are abelian,
D-branes on the orbifold were investigated in \cite{DGM,Muto1,KS,LNV}
and brane configurations corresponding to this case
are known as brane box models \cite{HZ,HU}.
For nonabelian cases,
gauge groups and field content were investigated in \cite{HH,GLR,Muto2}.
Brane configurations for D-type subgroups
$\Gamma=\Delta(3n^2)$ and $\Delta(6n^2)$
were also proposed in \cite{Muto3}
\footnote{Brane configurations for other types of nonabelian orbifolds
were discussed in \cite{FHH1,FHH2}.}.
The crucial point of the construction of the configuration is
the correspondence between quiver diagrams of $\Gamma$ and
brane configurations for $\C^3/\Gamma$.
According to the fact that
the quiver diagram of $\Delta(3n^2)$ ($\Delta(6n^2)$)
is a $\Z_3$ ($S_3$) quotient of the quiver diagram of
${\bf Z}_n \times {\bf Z}_n$,
the configuration for D-branes on ${\bf C}^3/\Delta(3n^2)$
(${\bf C}^3/\Delta(6n^2)$) is obtained from the configuration
for D-branes on ${\bf C}^3/{\bf Z}_n \times {\bf Z}_n$
by taking $\Z_3$ ($S_3$) quotient.
In order that the $\Z_3$ quotient can be defined,
brane configuration for $\C^3/\ZnZn$ must have
$\Z_3$ symmetry.
In \cite{Muto3}, the $\Z_3$ symmetry was realized by
a web of $(p,q)$ 5-branes.

It is interesting to study whether similar approach
can be applied to the cases in which
$\Gamma$ are exceptional type subgroups of $SU(3)$.
We can see that quiver diagrams of such subgroups
can be obtained from that of $\Delta(3 \times 3^2)$ by certain quotients:
for instance, the quiver diagram of an exceptional type
subgroup $\Sigma(648)$ is obtained by a quotient by the tetrahedral group.
From the correspondence between quiver diagrams and brane configurations,
we expect that a brane configuration for $\C^3/\Sigma(648)$
is a quotient of the configuration for $\C^3/\Delta(3 \times 3^2)$
by the tetrahedral group.
It is however difficult to realize such a quotient on the brane configuration
for $\C^3/\Delta(3 \times 3^2)$
since it has essentially two-dimensional structure
and it is not clear how the group acts on it.
The situation is similar to the case of the brane configuration
for $\C^3/\Delta(3n^2)$.
In that case,
we constructed a brane configuration for $\C^3/\ZnZn$
with a $\Z_3$ symmetry in order that we can define a $\Z_3$ quotient.
In the present case,
it would be natural that the brane configuration
for $\C^3/\Delta(3 \times 3^2)$
on which the tetrahedral group acts
has three-dimensional structure and
manifest symmetry under the tetrahedral group.

In this paper, 
as a step to the construction of brane configurations
corresponding to the E-type subgroups,
we propose brane configuraions for
${\bf C}^3/\Delta(3n^2)$ and ${\bf C}^3/\Delta(6n^2)$
with three-dimensional structure.
The point of the construction is to lift the quiver diagram
of $\ZnZn$, which has essentially two-dimensional structure,
into three-dimensions.
The configuration corresponding to this quiver diagram
consists of three kinds of NS5-branes intersecting each other
and D4-branes whose three directions are bounded by NS5-branes.
Naively, the configuration seems to be inappropriate since
the number of supersymmetries of the configuration
is half of the number the field theory should have:
the configuration have two supercharges
while field theories of D-branes on an orbifold $\C^3/\Gamma$
with $\Gamma \in SU(3)$ have four supercharges.
We argue, however, that
the supersymmetry of the field theory on D-branes is enhanced
and the number of supersymmetries is the same as that
of D-branes on $\C^3/\Gamma$ with $\Gamma \in SU(3)$.
By taking quotients on the configuration,
brane configurations for ${\bf C}^3/\Delta(3n^2)$
and ${\bf C}^3/\Delta(6n^2)$ can be constructed.

The organization of this paper is as follows.
In Section 2, we present three-dimensional realization
of the quiver diagrams of $\Gamma=\ZnZn, \Delta(3n^2)$ and $\Delta(6n^2)$.
In Section 3, we first review relations between
quiver diagrams and brane configurations.
Based on the relations,
we construct brane configuration
corresponding to D1-branes on orbifolds $\C^3/\ZnZn$,
$\C^3/\Delta(3n^2)$ and $\C^3/\Delta(6n^2)$.
We also argue the enhancement of supersymmetries of the field theory
on D-branes.
In Section 4, we comment on a possibility of constructing
configurations for $\C^3/\Gamma$
with $\Gamma$ an E-type subgroup of $SU(3)$
based on the configuration for $\C^3/\Delta(3 \times 3^2)$
constructed in Section 3.

\section{Three-dimensional realization of quiver diagrams}
\reseteqnum

In this section, we draw quiver diagrams of finite subgroups
of $SU(3)$ in three-dimensional form.
A quiver diagram of a finite group $\Gamma$,
which consists of nodes and arrows, represents
algebraic structure of irreducible representations of $\Gamma$,
\begin{equation}
R_3 \otimes R^a
= \oplus_b n_{ab} R^b.
\end{equation}
Here $R^a$ denotes irreducible representations of $\Gamma$
and $R_3$ is a faithful three-dimensional representation of $\Gamma$.
Each node of the quiver diagram
represents each irreducible representation of $\Gamma$
and the coefficient $n_{ab}$ is the number of arrows
starting from the $a$-th node and ending at the $b$-th node.

\subsection{Quiver diagrams of $\ZnZn$}

\begin{figure}[htdp]
\begin{center}
\leavevmode
\epsfysize=45mm
\epsfbox{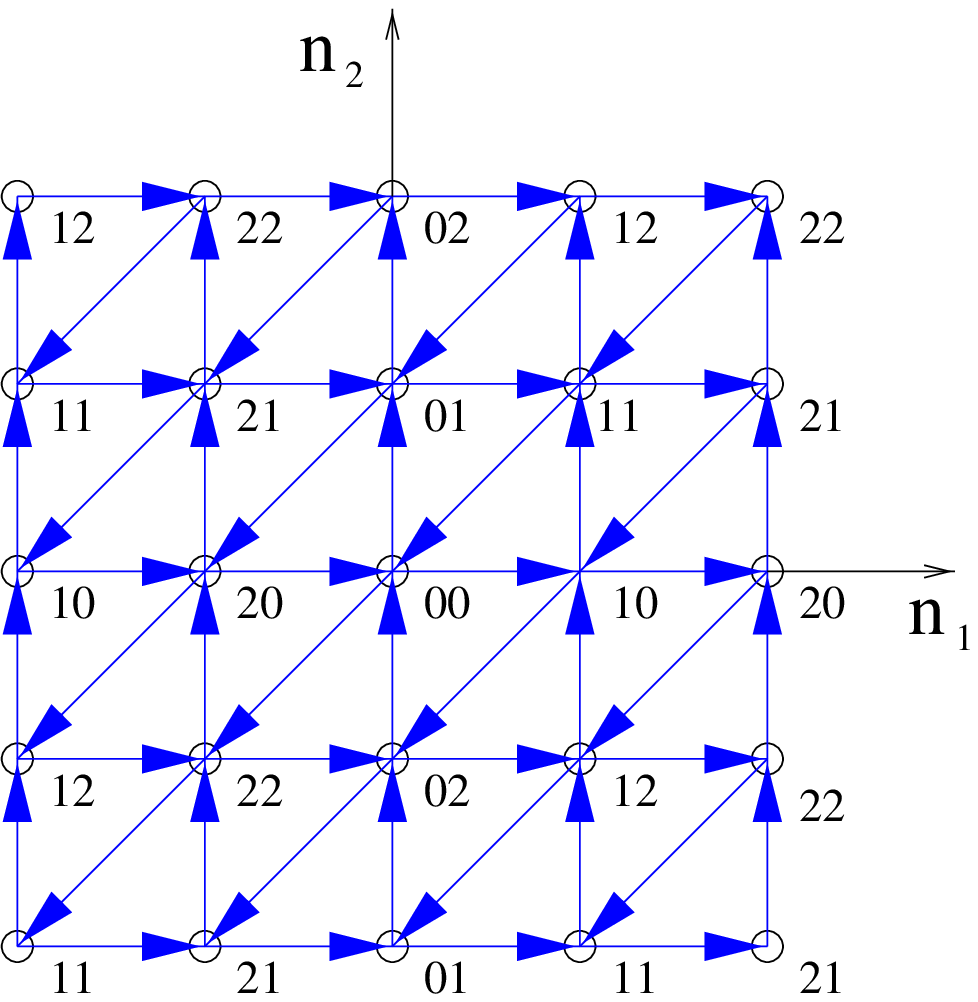}
\caption{The quiver diagram of $\ZnZn$ with $n=3$.}
\label{fig:quiver9}
\end{center}
\end{figure}

We first present the quiver diagram of an abelian subgroup
$\ZnZn$ of $SU(3)$.
Irreducible representations of ${\bf Z}_n \times {\bf Z}_n$
consist of $n^2$ one-dimensional representations $R_1^{(l_1,l_2)}$
with $(l_1,l_2) \in \ZnZn$.
We assign an irreducible representation $R_1^{(l_1,l_2)}$
to points of a two-dimensional lattice $\Z \times \Z$ represented as
\begin{equation}
(n_1,n_2)=(l_1+n\Z, l_2+n\Z).
\label{eq:assign}
\end{equation}
If we choose the three-dimensional representation to be
\begin{equation}
R_3 = R_1^{(1,0)}
\oplus R_1^{(0,1)}
\oplus R_1^{(-1,-1)},
\end{equation}
the decomposition of the product $R_3$ and $R_1^{(l_1,l_2)}$
becomes
\begin{equation}
R_3 \otimes R^{(l_1,l_2)}_1
= R^{(l_1+1,l_2)}_1
\oplus R^{(l_1,l_2+1)}_1
\oplus R^{(l_1-1,l_2-1)}_1.
\end{equation}
This equation implies that
there are three arrows which start from the node $(l_1,l_2)$;
the three end points are $(l_1+1,l_2)$, $(l_1,l_2+1)$ and $(l_1-1,l_2-1)$.
The quiver diagram of the group $\ZnZn$
is depicted in Figure \ref{fig:quiver9}.

\begin{figure}[htdp]
\begin{center}
\begin{minipage}{35mm}
\begin{center}
\leavevmode
\epsfysize=30mm
\epsfbox{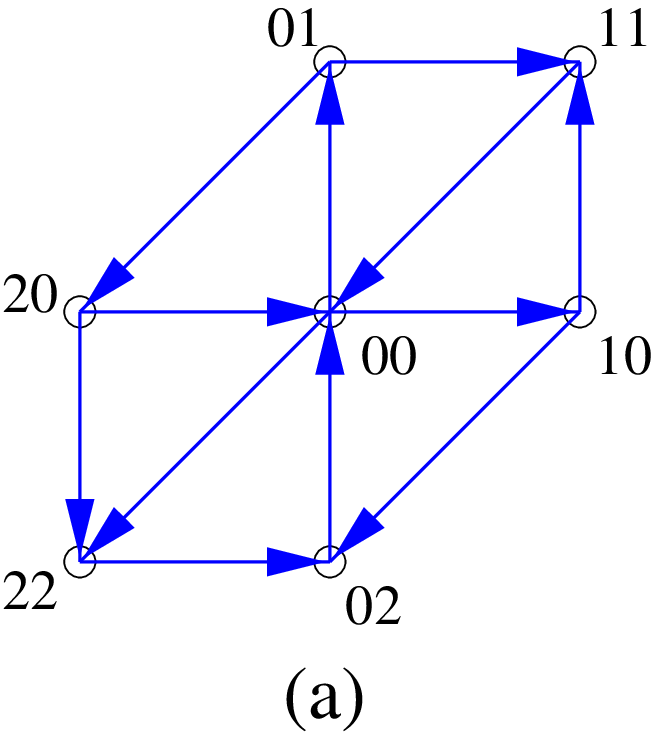}
\end{center}
\end{minipage}
\hspace{20mm}
\begin{minipage}{35mm}
\begin{center}
\leavevmode
\epsfysize=28mm
\epsfbox{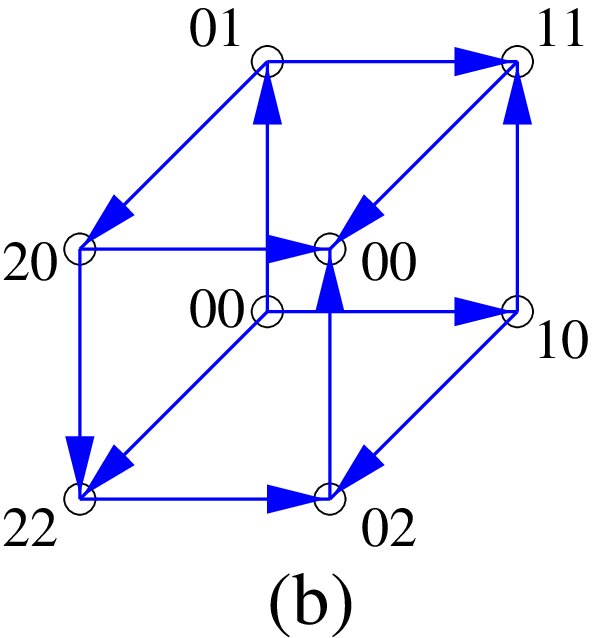}
\end{center}
\end{minipage}
\caption{
(a) A part of the quiver diagram of $\ZnZn$.
(b) A part of the quiver diagram of $\ZnZn$ in three-dimensional
form.
Figure \ref{fig:quiverunit}(a) can be considered as a projection of
Figure \ref{fig:quiverunit}(b) along a diagonal direction of the cube.}
\label{fig:quiverunit}
\end{center}
\end{figure}

Now we redraw the quiver diagram of Figure \ref{fig:quiver9}
in three-dimensional form.
The idea is that we consider the quiver diagram of Figure \ref{fig:quiver9}
as a projection of a certain quiver diagram in three-dimensional space
onto a certain plane.
Concretely speaking, we consider a part of the quiver diagram in
Figure \ref{fig:quiverunit}(a) as a projection of a cube in
Figure \ref{fig:quiverunit}(b) along a diagonal direction of the cube.

\begin{figure}[htdp]
\begin{center}
\leavevmode
\epsfysize=50mm
\epsfbox{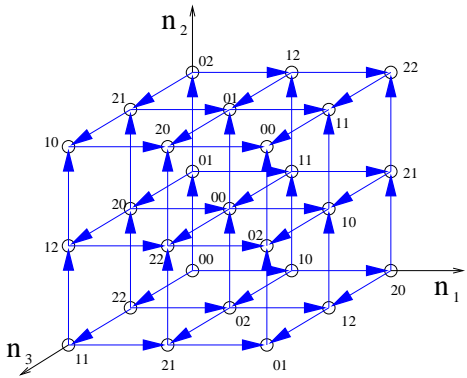}
\caption{Part of the quiver diagram of $\ZnZn$ with $n=3$
in three-dimensional form.}
\label{fig:quiver9-3}
\end{center}
\end{figure}

Under this consideration,
the assignment (\ref{eq:assign}) of the irreducible representations
to a two-dimensional lattice $\Z^2$
is translated into the assignment of $R_1^{(l_1,l_2)}$
to points on a three-dimensional lattice $\Z^3$ represented as
\begin{equation}
(n_1,n_2,n_3)=(l_1+n\Z+m, l_2+n\Z+m, m),
\label{eq:assign-3}
\end{equation}
where $m$ is an integer.
As indicated in Figure \ref{fig:quiverunit}(b),
three arrows going from the nodes $(n_1,n_2,n_3)$
have end points at $(n_1+1,n_2,n_3)$, $(n_1,n_2+1,n_3)$
and $(n_1,n_2,n_3+1)$.
An example of the quiver diagram is depicted in
Figure \ref{fig:quiver9-3}.

Note that the quiver diagram is uniform along the direction
$n_1=n_2=n_3$,
which becomes a key point in the discussion of
supersymmetry enhancement.

\subsection{Quiver diagrams of $\Delta (3n^2)$}

\begin{figure}[htdp]
\begin{center}
\begin{minipage}{60mm}
\begin{center}
\leavevmode
\epsfysize=50mm
\epsfbox{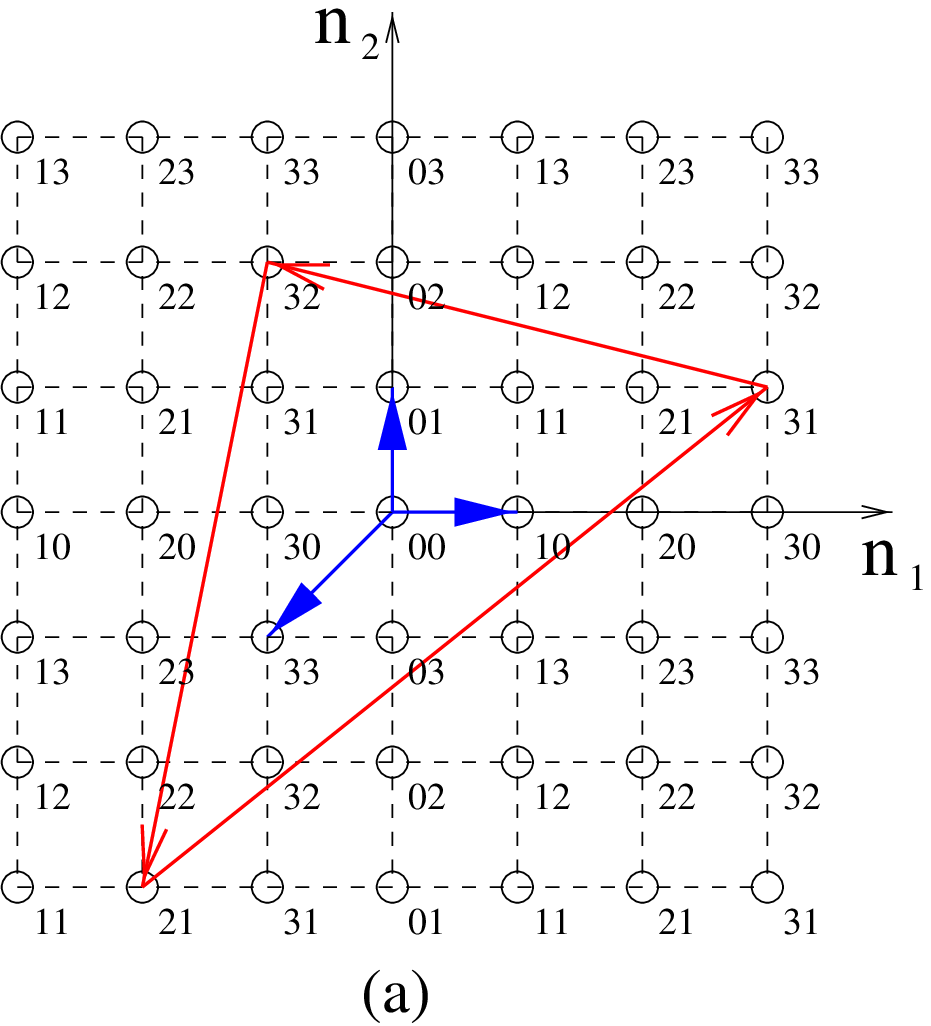}
\end{center}
\end{minipage}
\hspace{20mm}
\begin{minipage}{60mm}
\begin{center}
\leavevmode
\epsfysize=50mm
\epsfbox{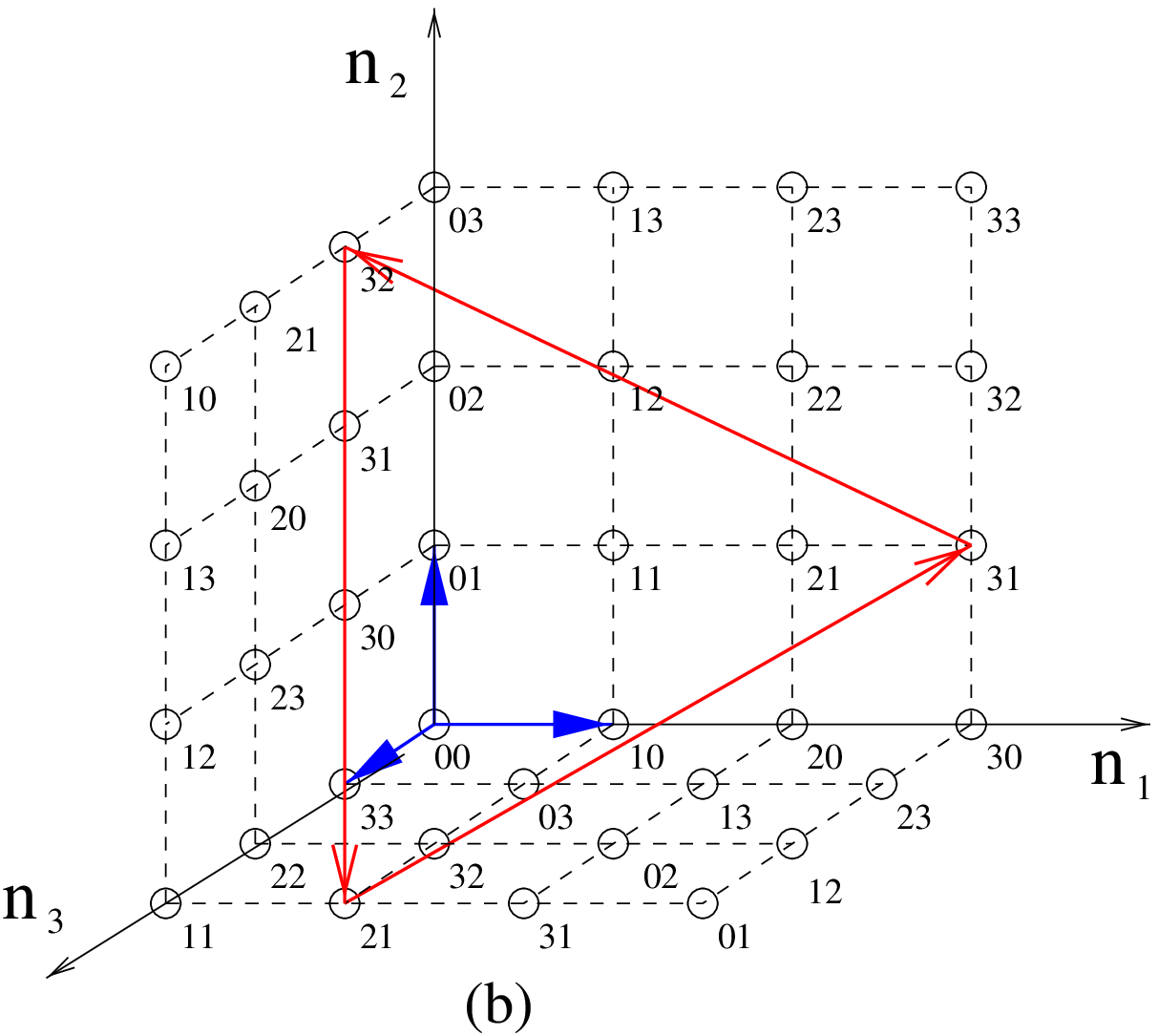}
\end{center}
\end{minipage}
\caption{
(a) The quiver diagram of $\Delta(3n^2)$ with $n=4$.
It is obtained from the quiver diagram of $\Z_4 \times \Z_4$
by the $\Z_3$ quotient.
Three nodes (3,1), (3,2) and (2,1) must be identified, for example.
Only one set of arrows is depicted for simplicity
although such a set of arrows start from every node.
(b) Three-dimensional version of
the quiver diagram of $\Delta(3n^2)$ with $n=4$.
Most of the nodes are omitted although
there is a node on every lattice point of $\Z^3$.}
\label{fig:quiver48}
\end{center}
\end{figure}

As discussed in \cite{Muto3},
the quiver diagram of the group $\Delta(3n^2)$
is obtained by a $\Z_3$ quotient of the quiver diagram of $\ZnZn$.
The $\Z_3$ acts on the label of irreducible representations as
\begin{equation}
(l_1,l_2) \rightarrow (-l_2,l_1-l_2) \rightarrow (-l_1+l_2,-l_1).
\label{eq:Z3}
\end{equation}
If the three nodes related by the $\Z_3$ action
correspond to different representations,
the $\Z_3$ quotient means that the three nodes
must be identified.
For instance,
the three nodes labeled by (3,1), (3,2) and (2,1)
must be identified if $n=4$.
Such nodes represent a three-dimensional
irreducible representation of $\Delta(3n^2)$.
For representations invariant under the $\Z_3$ action
such as $R_1^{(0,0)}$,
the $\Z_3$ quotient must be understood as a split of the node
into three nodes, each of which represents a one-dimensional
irreducible representation of $\Delta(3n^2)$.
The quiver diagram of the group $\Delta(3n^2)$ is depicted
in Figure \ref{fig:quiver48}(a).

We now consider a three-dimensional realization of the quiver diagram.
The idea is the same as the two-dimensional realization of the
quiver diagram.
That is, the quiver diagram of the group $\Delta(3n^2)$
is obtained by a $\Z_3$ quotient of the quiver diagram
of $\ZnZn$ given in Figure \ref{fig:quiver9-3}.
The action of $\Z_3$ on the lattice points of $\Z^3$
is defined as follows.
\begin{equation}
(n_1,n_2,n_3) \rightarrow (n_2,n_3,n_1) \rightarrow (n_3,n_1,n_2).
\label{eq:Z3-3}
\end{equation}
It means that $\Z_3$ acts on the three-dimensional space
as a $3\pi/2$ rotation along the line specified by $n_1=n_2=n_3$.
Combining with the assignment (\ref{eq:assign-3})
of the irreducible representations
of $\ZnZn$ on the lattice $\Z^3$,
one can see that the action (\ref{eq:Z3-3}) is equivalent to
the action of $\Z_3$ given in (\ref{eq:Z3}).
Three-dimensional version of the quiver diagram is depicted in
Figure \ref{fig:quiver48}(b).

\begin{figure}[htdp]
\begin{center}
\leavevmode
\epsfysize=45mm
\epsfbox{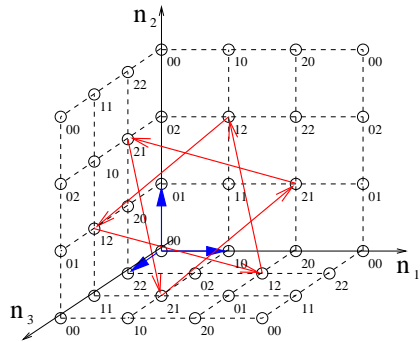}
\caption{Three dimensional realization of the quiver diagram
of $\Delta(3n^2)$ with $n=3$.
The nodes (2,1) and (1,2) become fixed points of the $Z_3$ action
in addition to (0,0).}
\label{fig:quiver27-3}
\end{center}
\end{figure}

Now we comment on the speciality when $n/3$ is an integer.
When $n/3$ is not an integer,
the only fixed node under the $\Z_3$ action is $(0,0)$.
On the other hand, when $n/3$ is an integer,
there are additional fixed nodes $(2n/3,n/3)$ and $(n/3,2n/3)$
as one can see from Figure \ref{fig:quiver27-3}.

\subsection{Quiver diagrams of $\C^3/\Delta (6n^2)$}

\begin{figure}[htdp]
\begin{center}
\begin{minipage}{60mm}
\begin{center}
\leavevmode
\epsfysize=50mm
\epsfbox{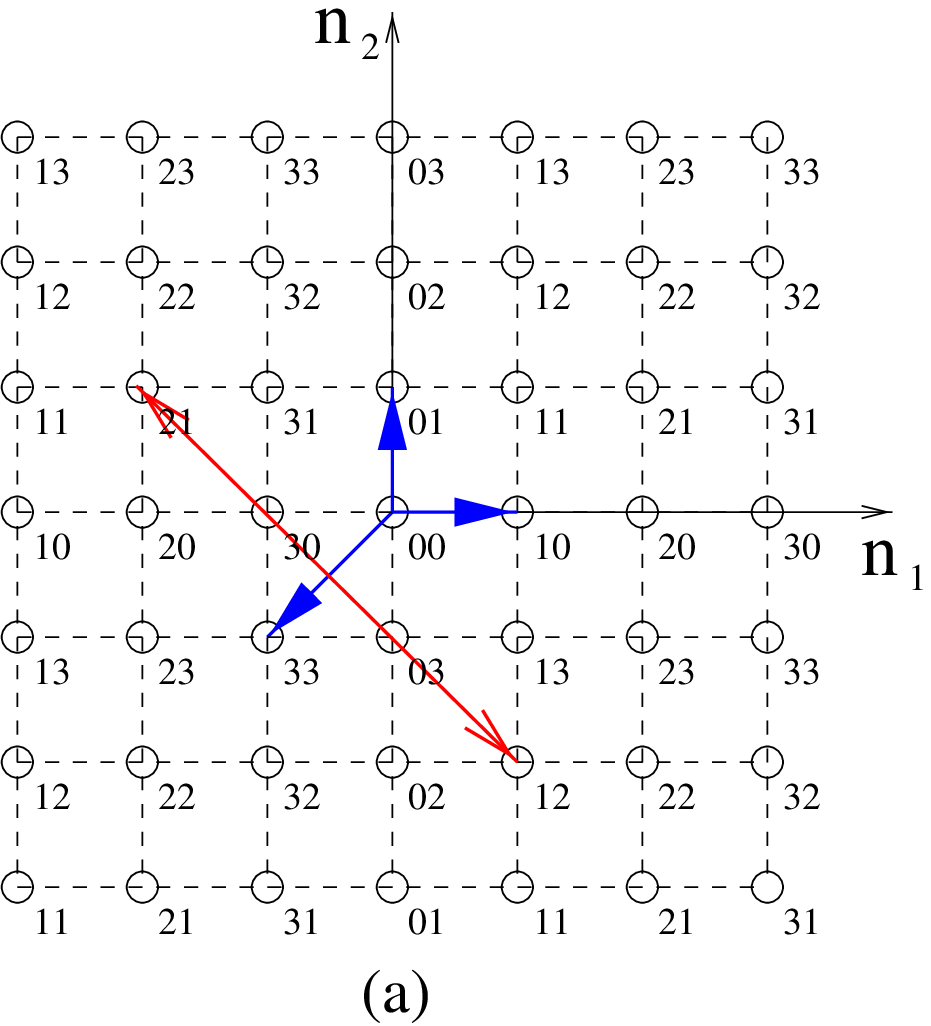}
\end{center}
\end{minipage}
\hspace{20mm}
\begin{minipage}{60mm}
\begin{center}
\leavevmode
\epsfysize=50mm
\epsfbox{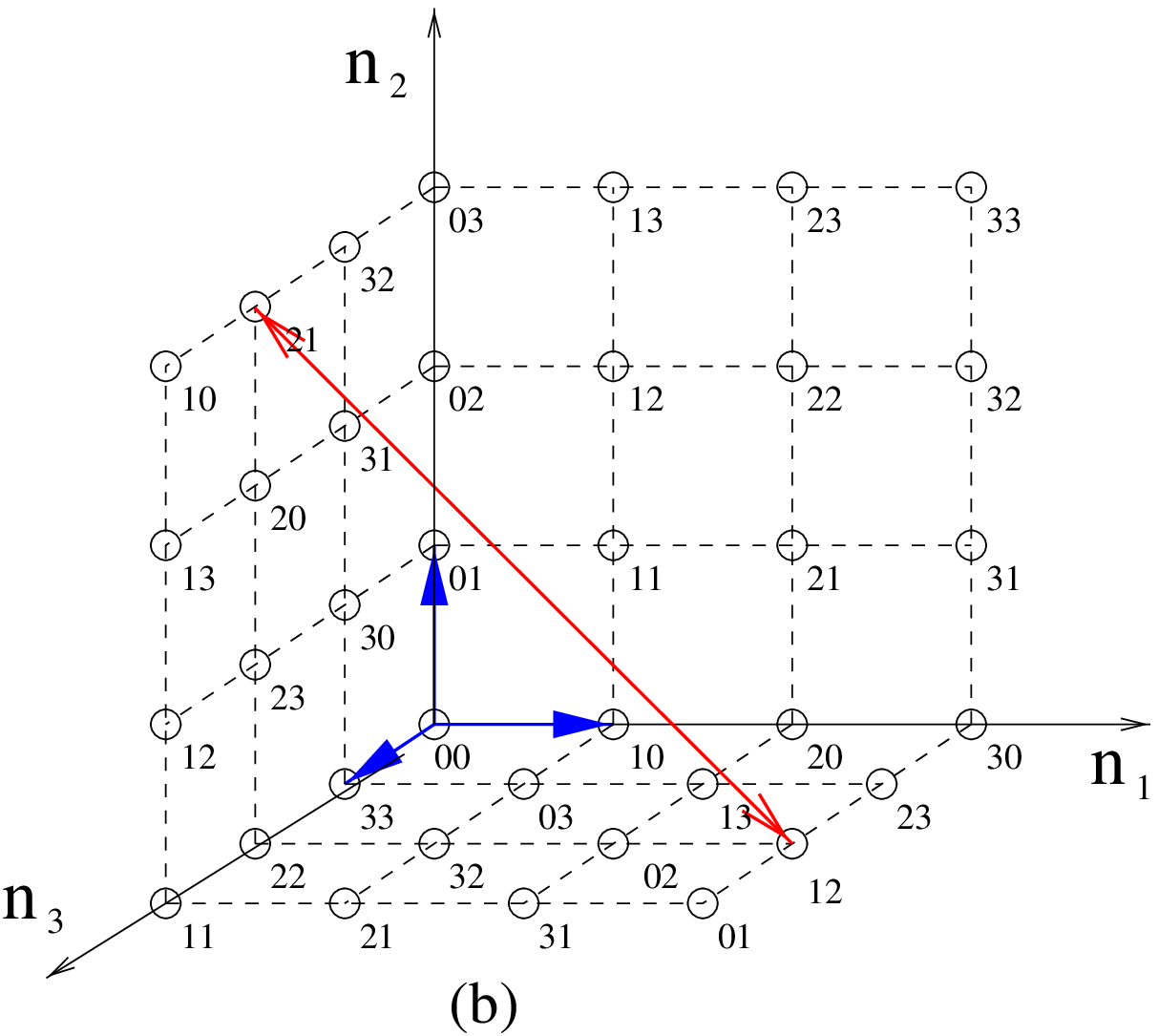}
\end{center}
\end{minipage}
\caption{
(a) The quiver diagram of $\Delta(6n^2)$ with $n=4$.
The nodes must be identified under the $\Z_2$ action
as well as the $\Z_3$ identification in the $\Delta(3n^2)$ case.
Two nodes (1,2) and (2,1) must be identified, for example.
(b) Three-dimensional version of the quiver diagram
of $\Delta(6n^2)$ with $n=4$.}
\label{fig:quiver96}
\end{center}
\end{figure}

The quiver diagram of the group $\Delta(6n^2)$
is obtained by a further $\Z_2$ quotient on the quiver diagram
of $\Delta(3n^2)$.
The $\Z_2$ acts on the label of irreducible representations as
\begin{equation}
(l_1,l_2) \rightarrow (l_2,l_1).
\label{eq:Z2}
\end{equation}
It means that $\Z_2$ acts as a reflection
with respect to the line extending a diagonal direction $n_1=n_2$.
If the two nodes related by the $\Z_2$ action
correspond to different representations,
they must be identified.
For instance,
the nodes (1,2) and (2,1), both of which are three-dimensional
irreducible representations of $\Delta(3n^2)$ for $n=4$,
must be identified.
Such nodes represent six-dimensional
irreducible representation of $\Delta(6n^2)$.
For nodes invariant under the $\Z_2$ action,
$\Z_2$ acts as a split the node into a certain set of nodes.
For details on the irreducible representations of $\Delta(6n^2)$,
see \cite{Muto3}.
The quiver diagram of the group $\Delta(6n^2)$ is depicted
in Figure \ref{fig:quiver96}(a).

In the three-dimensional version of the
quiver diagram,
$\Z_2$ acts on a three-dimensional lattice $\Z^3$ as
\begin{equation}
(n_1,n_2,n_3) \rightarrow (n_2,n_1,n_3).
\label{eq:Z2-3}
\end{equation}
It means that $\Z_2$ acts as a reflection 
with respect to the plane specified by $n_1=n_2$.
Combining with the $\Z_3$ action (\ref{eq:Z3-3}),
it is equivalent to the reflections with respect to the planes $n_2=n_3$
and $n_3=n_1$.
The three-dimensional version of the quiver diagram
of the group $\Delta(6n^2)$ is depicted
in Figure \ref{fig:quiver96}(b).

\section{Brane cube configurations for $\C^3/\Gamma$}
\reseteqnum

As discussed in \cite{HU,Muto3},
brane box type realization of D-brane gauge theory on an orbifold $\C^3/\Gamma$
has a direct correspondence to the quiver diagram of $\Gamma$.
In the D-brane gauge theory on the orbifold $\C^3/\Gamma$,
each node in the quiver diagram
corresponds to a gauge group factor $U(N_a)$
where $N_a$ is the dimension of the irreducible representation.
In the brane box type realization of the gauge theory,
it corresponds to a box with $N_a$ D-branes.
Arrows in the quiver diagram represent matter contents
of the D-brane gauge theory on $\C^3/\Gamma$.
That is, an arrow from the node $a$ to the node $b$ represents
a bifundamental matter transforming as $(N_a,{\bar N}_b)$ under
$U(N_a) \times U(N_b)$.
In the brane box type configuration,
it comes from an oriented open string starting from
D-branes on the $a$-th box and ending on D-branes
on the $b$-th box.
It implies that the $a$-th box and $b$-th box must adjoin each other.
These two boxes are separated by another brane, for example,
NS 5-branes or $(p,q)$ 5-branes.
Due to the orientation of such branes at the boundary,
only one orientation of open strings are allowed
and it induces a particular set of bifundamental matters \cite{HZ}.
Thus the arrows in the quiver diagram indicate
how to connect boxes with D-branes.
We summurize the correspondence in Table \ref{table:correspondence}.
\begin{table}[htdp]
\begin{center}
\caption{The correspondence among representation theory,
gauge theory, quiver diagram and brane box model.}
\label{table:correspondence}
\vspace{0.2cm}
\begin{tabular}{c|cc}
\hline
\hline
representation theory&$N_a$-dim irreducible repr.
&$R_3 \otimes R^a=\oplus n_{ab}R^b$ \\
gauge theory&$U(N_a)$&matter $(N_a, \bar{N_b})$\\
quiver diagram&node $a$&arrow from $a$ to $b$\\
brane box model&box with $N_a$ D-branes
&oriented open strings\\ \hline
\end{tabular}
\end{center}
\end{table}

It is also important that the configuration
provides the same supersymmetry as that of the
D-brane gauge theory on the orbifold.
Several brane configurations satisfying these reqirement
were constructed.
In \cite{HZ,HU}, a brane configuration is constructed
for $\Gamma=\Z_n \times \Z_m$ by using
D5-branes and two kinds of NS5-branes.
In \cite{Muto3}, brane configurations were constructed for
the nonabelian groups $\Delta(3n^2)$ and $\Delta(6n^2)$
as well as the abelian group $\ZnZn$
using $(p,q)$ 5-branes and D3-branes.

In this section, we construct another kind of brane
configurations for $\Gamma=\ZnZn$, $\Delta(3n^2)$ and $\Delta(6n^2)$
based on the three-dimensional version of the quiver diagrams
given in the last section.

\subsection{Brane cube configurations for $\C^3/\ZnZn$}

\begin{figure}[htdp]
\begin{center}
\leavevmode
\epsfysize=35mm
\epsfbox{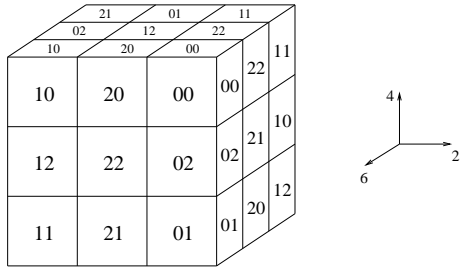}
\caption{The brane cube configuration for $\C^3/\ZnZn$ with $n=3$.
The cube at $(n_1,n_2,n_3)$ has the same label
as the cube at $(n_1+1,n_2+1,n_3+1)$.}
\label{fig:brane9-3a}
\end{center}
\end{figure}

In this subsection, we construct brane configurations for $\C^3/\ZnZn$
based on the three-dimensional version of the quiver
diagram given in Figure \ref{fig:quiver9-3}.
As the nodes in the quiver diagram lie at the lattice points of $\Z^3$,
boxes also lie at the lattice points of $\Z^3$.
The node at $(n_1,n_2,n_3)$ is connected
to the nodes at $(n_1+1,n_2,n_3)$,
$(n_1,n_2+1,n_3)$ and $(n_1,n_2,n_3+1)$ in the quiver diagram
by three outgoing arrows,
so the box at $(n_1,n_2,n_3)$ must adjoin boxes at $(n_1+1,n_2,n_3)$,
$(n_1,n_2+1,n_3)$ and $(n_1,n_2,n_3+1)$.
A natural brane configuration satisfying these requirement
is depicted in Figure \ref{fig:brane9-3a}.
Note that the cube at $(n_1,n_2,n_3)$ has the same label
as the cube at $(n_1+1,n_2+1,n_3+1)$.

The configuration consists of the following branes:
\begin{itemize}
\item NS5-branes located along 012345 directions.
\item NS'5-branes located along 012367 directions.
\item NS''5-branes located along 014567 directions.
\item D4-branes located along 01246 directions.
\end{itemize}

D4-branes are bounded in the direction 6 by the NS5-branes,
in the direction 4 by the NS'5-branes,
and in the direction 2 by the NS''5-branes.
Thus the non-compact directions of the D4-branes are 0 and 1,
and hence the low energy theory becomes two-dimensional.
The numbers written in the boxes are the labels
of the irreducible representations of $\ZnZn$.
Due to the orientation of NS5-branes,
only one orientation of open strings is allowed
and it gives a bifundamental matter
corresponding to the arrow of the quiver diagram.

There is however a subtlety on the number of supersymmetries.
The configuration consists of four kinds of branes,
each of which breaks 1/2 supersymmetries.
Thus the brane cube configuration has two supercharges.
Since the number of supersymmetries of the gauge theory
of D-branes on $\C^3/\Gamma$ with $\Gamma \in SU(3)$ is four,
it seems that the two gauge theories have different numbers of
supersymmetries.
As we will discuss, however,
the low energy field theory of D-branes obtained from
the brane cube configuration has an enough number of supersymmetries
due to a certain enhancement of supersymmetry.

To explain the supersymmetry enhancement,
we derive the brane cube configuration from different point of view.
It is obtained from D1-branes on $\C^4/\Gamma$ by performing
T-duality three times.
In fact, a similar but different configuration
were discussed in \cite{GU,KLM}.
Both of them have a property that the configuration
is uniform along one direction.
It is the key point of the supersymmetry enhancement.

We start with the D1-brane gauge theory
on $\C^4/\Gamma$ with $\Gamma$ an abelian
subgroup of $SU(4)$ \cite{Mohri1,GU}.
We consider $|\Gamma|$ D1-branes on $\C^4$
and let the D1-branes extend along 01 directions.
The low energy effective action of the D1-brane
is given by the dimensional reduction of ten-dimensional
${\cal N}=1$ supersymmetric $U(|\Gamma|)$ Yang-Mills theory
to two dimensions.
The theory contains gauge field $A$,
four complex scalar fields $Z^\mu$ ($\mu=1,2,3,4$),
eight left-handed fermions $\lambda_-$
and eight right-handed fermions $\psi_+$.
These fields transforms in the adjoint of $U(|\Gamma|)$.
Then we project the theory onto $\Gamma$ invariant states.
To perform the projection,
one must define how $\Gamma$ acts on $\C^4$.
This is specified by a four-dimensional representation $R_4$.
We denote the decomposition of $R_4$
into irreducible representations
$R^a$ ($a=1,\dots,|\Gamma|$) of $\Gamma$ as
\begin{equation}
R_4 = R^{a_1} \oplus R^{a_2} \oplus R^{a_3} \oplus R^{a_4}.
\end{equation}
(As $\Gamma$ is abelian, its irreducible representations are
one-dimensional.)
Each irreducible representation $R^{a_\mu}$
acts on the coordinates $z^\mu$ of $\C^4$
as $z^\mu \rightarrow R^{a_\mu} z^\mu$.
The requirement that $\Gamma$ is a subgroup of $SU(4)$ can be stated as
\begin{equation}
R^{a_1} \otimes R^{a_2} \otimes R^{a_3} \otimes R^{a_4} = R^0
\label{eq:SU(4)}
\end{equation}
where $R^0$ is the trivial representation.
If we write $R^a \otimes R^b = R^{a \oplus b}$,
the condition is represented as
$R^{a_4}=R^{-a_1-a_2-a_3}$.
To define D-branes on $\C^4/\Gamma$,
one must also determine how $\Gamma$ acts on the Chan-Paton indices.
We define the action of $\Gamma$ on the Chan-Paton indices
as the regular representation of $\Gamma$.
Note that the theory obtained by the projection
is a two-dimensional (0,2) supersymmetric theory with the gauge group
$U(1)^{|\Gamma|}=\Pi_a U(1)_a$,
where $U(1)_a$ represents the gauge group $U(1)$
corresponding to the representation $R^a$.

The field content surviving the projection is as follows:
\begin{itemize}
\item a gauge field $A_a$ of $U(1)_a$,
\item four complex bosons $Z_a^\mu$
which transform in the $(\fund, \overline{\fund})$ of
$U(1)_a \times U(1)_{a \oplus a_\mu}$,
\item a left-handed Dirac fermion $\lambda_{-a}$
which transform in the adjoint of $U(1)_a$,
\item three left-handed Dirac fermions $\lambda_{-a}^{\mu\nu}$
which transforms in the $(\fund, \overline{\fund})$ of
$U(1)_a \times U(1)_{a \oplus a_\mu \oplus a_\nu}$,
\item four right-handed Dirac fermions $\psi_{+a}^\mu$
which transforms in the $(\fund, \overline{\fund})$ of
$U(1)_a \times U(1)_{a \oplus a_\mu}$,
\end{itemize}
where $a$ runs from 1 to $|\Gamma|$.
Here $\lambda_{-a}^{\mu\nu}$ is antisymmetric in $\mu\nu$,
so there are six fields to each $a$,
three of which are independent.
There is an ambiguity to determine the three independent left-handed Dirac
fermions.
We take three fields $\lambda_{-a}^{i4}$ ($i=1,2,3$) be independent.

These fields form (0,2) multiplets as follows
\footnote{For details on supersymmetric field theories in two-dimensions,
see \cite{Witten}.}:
\begin{itemize}
\item The gauge field $A_a$ and the left-handed fermion $\lambda_{-a}$
and an auxiliary field $D_a$ form a (0,2) gauge multiplet $V_a$,
\begin{equation}
V_a=A_{0a}-A_{1a}-2i\theta^+ {\bar \lambda}_{-a}
-2i{\bar \theta}^+ \lambda_{-a} +2 \theta^+ {\bar \theta}^+ D_a,
\label{eq:02gaugem}
\end{equation}
\item Four complex bosons $Z_a^\mu$ and four right-handed fermions
$\psi_{+a}^\mu$ form four (0,2) chiral multiplets $\Phi_a^\mu$,
\begin{equation}
\Phi_a^\mu=Z_a^\mu+\sqrt 2 \theta^+ \psi_{+a}^\mu
-i \theta^+ {\bar \theta}^+ Z_a^\mu,
\label{eq:02chiralm}
\end{equation}
\item Three left-handed fermions $\lambda_{-a}^{i4}$
and auxiliary fields $G_a^i$
form three (0,2) Fermi multiplets $\Lambda_a^i$,
\begin{equation}
\Lambda_a^i=\lambda_{-a}^{i4} -\sqrt 2 \theta^+ G_a^i
-i \theta^+ {\bar \theta}^+
(D_0+D_1) \lambda_{-a}^{i4} -\sqrt 2 {\bar \theta}^+ E_a^i
\label{eq:02Fermim}
\end{equation}
where $D_\alpha$ is the supersymmetric derivative
and $E_a$ is a function of chiral superfields which will be
defined below.
\end{itemize}

The two-dimensional (0,2) supersymmetric gauge theories
are described by a Lagrangian of the form,
\begin{equation}
L=L_{gauge}+L_{ch}+L_F+L_J+L_{D,\theta},
\label{eq:02total}
\end{equation}
\begin{equation}
L_{gauge}=\frac{1}{8e_a^2} \int d^2 x d \theta^+ d {\bar \theta}^+
{\bar \Upsilon}_a \Upsilon_a,
\label{eq:02gauge}
\end{equation}
\begin{equation}
L_{ch}=-\frac{i}{2} \int d^2 x d \theta^+ d {\bar \theta}^+
{\bar \Phi}_a^\mu ({\cal D}_0-{\cal D}_1)\Phi_a^\mu,
\label{eq:02ch}
\end{equation}
\begin{equation}
L_F=-\frac{1}{2} \int d^2 x d \theta^+ d {\bar \theta}^+
{\bar \Lambda}_a^i \Lambda_a^i,
\label{eq:02F}
\end{equation}
\begin{equation}
L_J=-\frac{1}{\sqrt 2} \int d^2 x d \theta^+
\Lambda_a^i J_a^i |_{{\bar \theta}^+=0}-h.c.,
\label{eq:02J}
\end{equation}
\begin{equation}
L_{D,\theta}=\frac{t_a}{4} \int d^2 x d \theta^+
\Upsilon_a |_{{\bar \theta}^+=0} + h.c.,
\label{eq:02D}
\end{equation}
where $\Upsilon_a$ is the field strength of the superspace
gauge field $V_a$,
${\cal D}_\alpha$ is the gauge covariant derivative
and $t_a=\theta_a/2\pi+ir_a$ represents
a Fayet-Iliopoulos parameter and a theta parameter of $U(1)_a$.
The interactions of the theory are completely defined by functions
$J_a^i$ and $E_a^i$.
In the present case,
$J_a^i$ and $E_a^i$ take the following form,
\begin{equation}
J_a^i = \epsilon_{ijk}
\Phi^j_{a \oplus a_i}
\Phi^k_{a \oplus a_i \oplus a_j},
\label{eq:J}
\end{equation}
\begin{equation}
E_a^i = \Phi^4_a\Phi^i_{a \oplus a_4}
-\Phi^i_a\Phi^4_{a \oplus a_i}.
\label{eq:E}
\end{equation}
They satisfy the following relation
\begin{equation}
\sum_a \sum_i J^i_a E^i_a=0.
\label{eq:JE}
\end{equation}

\begin{figure}[htdp]
\begin{center}
\leavevmode
\epsfysize=40mm
\epsfbox{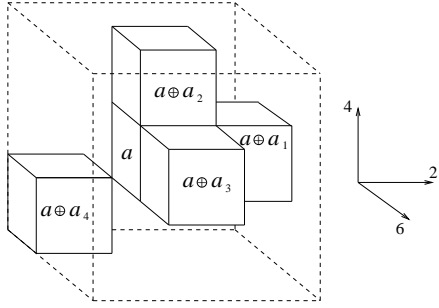}
\caption{The brane cube configuration obtained from $\C^4/\Gamma$
by T-dualities along compact directions associated with
$z_1$, $z_2$ and $z_3$.
Only a few boxes are depicted for clarity.
The label on each cube represents an irreducible representation
corresponding to the cube.}
\label{fig:branec4a}
\end{center}
\end{figure}

\begin{figure}[ht]
\begin{center}
\leavevmode
\epsfysize=35mm
\epsfbox{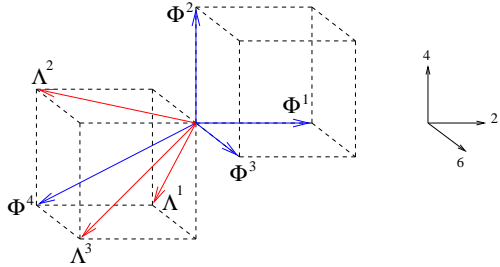}
\caption{The arrows representing various (0,2) multiplets
corresponding to the configuration of Figure \ref{fig:branec4a}.
$\Phi^\mu$ represents a chiral multiplet and $\Lambda^i$
represents a Fermi multiplet.}
\label{fig:quiverc4a}
\end{center}
\end{figure}

To realize the gauge theory in terms of brane configurations,
we perform T-dualities
three times along $U(1)$ orbits associated with
three complex planes
\footnote{To be precise, we must substitute $\C^4/\Gamma$
by a manifold with the same singularity
but different asymptotics to make the radius of
the $U(1)$ orbits finite at infinity.}.
There is an ambiguity to choose three directions
out of four complex coordinates.
If we T-dualize along three compact directions
associated with $z^1$, $z^2$ and $z^3$,
we obtain a brane cube configuration depicted in
Figure \ref{fig:branec4a}.
It consists of three kinds of NS5-branes and D4-branes.
The (0,2) chiral multiplet $\Phi_a^\mu$ transforming in the
$(\fund_a,\overline{\fund}_{a \oplus a_\mu})$ comes from open strings
extending from the box labeled by $a$ to the box labeled by $a \oplus a_\mu$.
We can represent the four chiral multiplets
$\Phi_a^1$, $\Phi_a^2$, $\Phi_a^3$ and $\Phi_a^4$
by arrows with components $(1,0,0)$, $(0,1,0)$, $(0,0,1)$ and $(-1,-1,-1)$.
The (0,2) Fermi multiplet ${\bf \Lambda}_a^i$ transforming in the
$(\fund_a,\overline{\fund}_{a \oplus a_i \oplus a_4})$ comes from open strings
from the box $a$ to the box $a \oplus a_i \oplus a_4$.
We can represent the three Fermi multiplets ${\bf \Lambda}_a^1$,
${\bf \Lambda}_a^2$ and ${\bf \Lambda}_a^3$
by arrows with components $(0,-1,-1)$, $(-1,0,-1)$ and $(-1,-1,0)$.
The arrows are depicted in Figure \ref{fig:quiverc4a}.

Now we restrict the model to the cases with $R^{a_4}=R^0$.
Then the condition for $\Gamma \in SU(4)$
in (\ref{eq:SU(4)}) becomes
\begin{equation}
R^{a_1} \otimes R^{a_2} \otimes R^{a_3} = R^0.
\label{eq:SU(3)}
\end{equation}
It means that
the orbifold $\C^4/\Gamma$ is
$\C^3/\Gamma \times \C$ where $\Gamma$ is a subgroup
of $SU(3)$.
In this case,
the supersymmetry of the two-dimensional theory
is (2,2) instead of (0,2),
and the (0,2) multiplets are combined into (2,2) multiplets
as follows:
\begin{itemize}
\item The (0,2) gauge multiplet $V_a$ and the adjoint
(0,2) chiral multiplet $\Phi_a^4$ are combined
to form (2,2) vector multiplet $V'_a$.
($\Phi_a^4$ transforms in the adjoint of $U(1)_a$ due to $R^{a_4}=R^0$.)
\begin{eqnarray}
&&V'_a = \theta^- {\bar \theta}^- (A_{0a}-A_{1a})
+\theta^+ {\bar \theta}^+ (A_{0a}+A_{1a})
-\sqrt 2 \theta^- {\bar \theta}^+ Z_a^4
-\sqrt 2 \theta^+ {\bar \theta}^- {\bar Z}_a^4 \nonumber \\
&&+2i \theta^- \theta^+
({\bar \theta}^- {\bar \lambda}_{-a} + {\bar \theta}^+ \psi_{+a}^4)
-2i {\bar \theta}^- {\bar \theta}^+
(\theta^- \lambda_{-a} + \theta^+ {\bar \psi}_{+a}^4)
-2\theta^- \theta^+ {\bar \theta}^- {\bar \theta}^+ D_a
\label{eq:22vectorm}
\end{eqnarray}
\item The (0,2) chiral multiplets $\Phi_a^i$
and the (0,2) Fermi multiplets $\Lambda_a^i$
form three (2,2) chiral multiplets $\Phi_a^{'i}$.
($\Lambda_a^i$ transforms in the
$(\fund_a,\overline{\fund}_{a \oplus a_i})$ due to $R^{a_4}=R^0$.)
\begin{equation}
\Phi_a^{'i}=Z_a^i(y) +\sqrt 2 \theta^- \lambda_{-a}^{i4}(y)
+\sqrt 2 \theta^+ \psi_{+a}^i(y)
-2\theta^- \theta^+ G_a^i(y)
\label{eq:22chiralm}
\end{equation}
where $y^0= x^0 -i(\theta^+ {\bar \theta}^+ + \theta^- {\bar \theta}^-)$
and $y^1= x^1 -i(\theta^+ {\bar \theta}^+ - \theta^- {\bar \theta}^-)$.
\end{itemize}

In the brane cube configuration in Figure \ref{fig:branec4a},
the cube at $(n_1,n_2,n_3)$
and the cube at $(n_1+1,n_2+1,n_3+1)$ are equivalent
due to the condition $R^{a_4}=R^0$.
This is nothing but the configuration
considered in Figure \ref{fig:brane9-3a}
if we take $\Gamma=\Z_3 \times \Z_3$,
$R^{a_1}=R_1^{(1,0)}$,
$R^{a_2}=R_1^{(0,1)}$ and
$R^{a_3}=R_1^{(-1,-1)}$.
The (2,2) chiral multiplets $\Phi_a^1$, $\Phi_a^2$ and $\Phi_a^3$
can be represented by arrows with components
$(1,0,0)$, $(0,1,0)$ and $(0,0,1)$.
The (2,2) vector multiplet $V_a$ can be represented by an arrow
with components $(-1,-1,-1)$.

The actions $L_{gauge}$ (\ref{eq:02gauge})
and $\mu=4$ part of $L_{ch}$ (\ref{eq:02ch})
of the (0,2) supersymmetric theory are combined into
$L_{gauge}$ of the (2,2) supersymmetric theory
\begin{equation}
L_{gauge}=-\frac{1}{4e_a^2} \int d^2 x d^4 \theta
{\bar \Sigma}_a \Sigma_a
\label{eq:22gauge}
\end{equation}
with some interaction terms.
Here $\Sigma_a$ is the field strength of the (2,2) superspace
gauge field $V'_a$.
Remaining part of $L_{ch}$ (\ref{eq:02ch}) and $L_F$ (\ref{eq:02F})
become $L_{ch}$ of the (2,2) supersymmetric theory,
\begin{equation}
L_{ch}= \int d^2 x d^4 \theta
{\bar \Phi}_a^{'i} e^{V'} \Phi_a^{'i}.
\label{eq:22ch}
\end{equation}
$L_J$ (\ref{eq:02J}) corresponds to the superpotential term $L_W$
with $W=\epsilon_{ijk} \Phi_a^{'i} \Phi_{a \oplus a_i}^{'j}
\Phi_{a \oplus a_i \oplus a_j}^{'k}$,
\begin{equation}
L_W=-\int d^2 x d \theta^+ d \theta^-
W(\Phi_a^{'i}) |_{{\bar \theta}^+={\bar \theta}^-=0}-h.c..
\label{eq:22W}
\end{equation}
Three arrows representing the three chiral multiplets
appearing in the superpotential form a triangle
up to the identification along the diagonal direction.
Note that the superpotential $W(\Phi_a^{'i})$ is related to $J_a^i$ as
\begin{equation}
J_a^i=\frac{\partial W}{\partial \Phi_a^{'i}},
\label{eq:JW}
\end{equation}
and the equation (\ref{eq:JE}) implies the gauge invariance
of the superpotential.
Combining with the D-term part
\begin{equation}
L_{D,\theta}=\frac{t_a}{4} \int d^2 x d \theta^+ d {\bar \theta}^-
\Sigma_a |_{\theta^-={\bar \theta}^+=0} + h.c.,
\label{eq:22D}
\end{equation}
one obtain two-dimensional (2,2) supersymmetric gauge theory.
Thus although the brane cube configuration has only two supercharges
the two-dimensional field theory on D-branes has an
enough number of supersymmetries.

\begin{figure}[htdp]
\begin{center}
\leavevmode
\epsfysize=40mm
\epsfbox{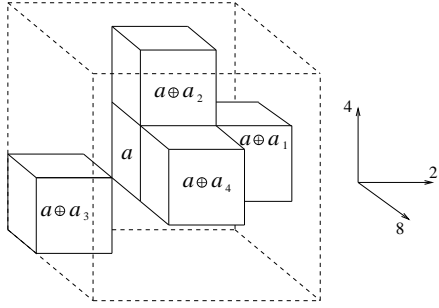}
\caption{The brane cube configuration obtained from $\C^4/\Gamma$
by T-dualities along compact directions associated with
$z_1$, $z_2$ and $z_4$.
The label on each cube represents an irreducible representation
corresponding to the cube.}
\label{fig:branec4b}
\end{center}
\end{figure}

To understand the reason for the supersymmetry enhancement,
it is useful to compare the configuration of Figure \ref{fig:brane9-3a}
with a brane configuration obtained by another T-duality.
As noted earlier, there is an ambiguity to choose three directions
to perform T-duality out of four complex coordinates.
If we T-dualize along three compact directions
associated with $z^1$, $z^2$ and $z^4$,
we obtain a brane cube configuration depicted in
Figure \ref{fig:branec4b}.

\begin{figure}[htdp]
\begin{center}
\leavevmode
\epsfysize=35mm
\epsfbox{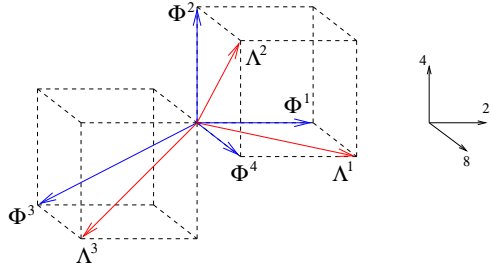}
\caption{The arrows representing various (0,2) multiplets
corresponding to the configuration of Figure \ref{fig:branec4b}.}
\label{fig:quiverc4b}
\end{center}
\end{figure}

The chiral multiplet $\Phi_a^\mu$ transforming in the
$(\fund_a,\overline{\fund}_{a \oplus a_\mu})$ comes from open strings
from the box $a$ to the box $a \oplus a_\mu$.
We can represent the four chiral multiplets
$\Phi_a^1$, $\Phi_a^2$, $\Phi_a^3$ and $\Phi_a^4$
by arrows with components $(1,0,0)$, $(0,1,0)$, $(-1,-1,-1)$ and $(0,0,1)$.
The Fermi multiplet ${\bf \Lambda}_a^i$ transforming in the
$(\fund_a,\overline{\fund}_{a \oplus a_i \oplus a_4})$ comes from open strings
from the box $a$ to the box $a \oplus a_i \oplus a_4$.
We can represent the three Fermi multiplets ${\bf \Lambda}_a^1$,
${\bf \Lambda}_a^2$ and ${\bf \Lambda}_a^3$
by arrows with components $(1,1,0)$, $(0,1,1)$ and $(-1,-1,0)$.
The arrows are depicted in Figure \ref{fig:quiverc4b}.

We would like to emphasize that the two configurations
in Figure \ref{fig:branec4a} and Figure \ref{fig:branec4b}
give the same field theory
since both of them are related to the D-branes on $\C^4/\Gamma$
by T-dualities
and hence the two configurations are related by T-duality.

\begin{figure}[htdp]
\begin{center}
\leavevmode
\epsfysize=35mm
\epsfbox{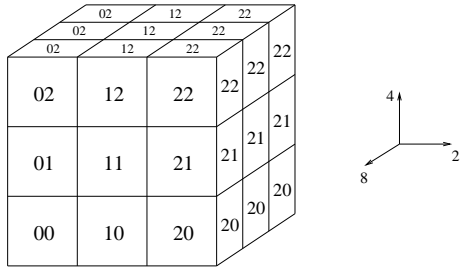}
\caption{The brane cube configuration for $\C^3/\ZnZn$ with $n=3$
obtained by T-dualities along 2,4,8 directions.
It is uniform along the direction 8.}
\label{fig:brane9-3b}
\end{center}
\end{figure}

In the brane cube configuration given in Figure \ref{fig:branec4b},
the cube at $(n_1,n_2,n_3)$
and the cube at $(n_1,n_2,n_3+1)$ are equivalent.
If we set $R^{a_4}=R^0$,
the (2,2) chiral multiplets $\Phi_a^1$, $\Phi_a^2$ and $\Phi_a^3$
can be represented by arrows with components
$(1,0,0)$, $(0,1,0)$ and $(-1,-1,-1)$,
while the (2,2) vector multiplet $V_a$ can be represented by an arrow
with components $(0,0,1)$.
This is the model considered in Section 3.2 of \cite{GU}.
That is, if we take $\Gamma=\Z_3 \times \Z_3$,
$R^{a_1}=R_1^{(1,0)}$,
$R^{a_2}=R_1^{(0,1)}$ and
$R^{a_3}=R_1^{(-1,-1)}$,
we obtain the configuration in Figure \ref{fig:brane9-3b}.

In the configuration of Figure \ref{fig:brane9-3b},
we can remove NS''5-branes extending along 012345 directions
without changing the matter contents and gauge groups
since the configuration of Figure \ref{fig:brane9-3b}
is trivial along the direction 8.
The resulting configuration is equivalent (up to a certain T-duality)
to the usual brane box model \cite{HZ,HU} with two kinds of NS5-branes.
Therefore the field theory obtained from the configuration
in Figure \ref{fig:brane9-3b} is equivalent to the field theory
obtained from the usual brane box model,
so the field theory has four supercharges.
As stated above, the configuration of Figure \ref{fig:brane9-3a}
is related to Figure \ref{fig:brane9-3b} T-duality,
the field theory realized by the configuration
has the enough number of supersymmetries.

Note that the configuration of Figure \ref{fig:brane9-3a}
is more natural than that of Figure \ref{fig:brane9-3b}
in the sense that the three coordinates of $\C^3$
(three chiral multiplets $\Phi^{'i}$) are treated equivalently.
This is the key point of the construction of the configuration
for $\C^3/\Delta(3n^2)$.
Due to the symmetric assignment of the multiplets, however,
enhancement of supersymmetry becomes rather nontrivial.

\subsection{Brane cube configurations for $\C^3/\Delta(3n^2)$}

\begin{figure}[htdp]
\begin{center}
\leavevmode
\epsfysize=40mm
\epsfbox{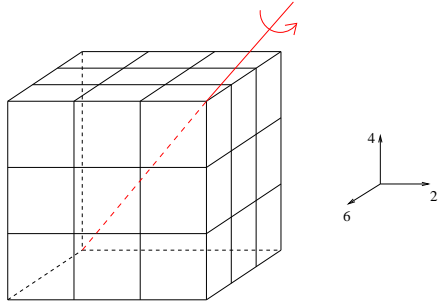}
\caption{The brane cube configuration for $\C^3/\Delta(3n^2)$ with $n=3$.
It is obtained from the brane cube configuration
in Figure \ref{fig:brane9-3a} by a $\Z_3$ quotient.
$\Z_3$ acts as a $2\pi/3$ rotation along the line
$x^2=x^4=x^6$.}
\label{fig:brane27-3}
\end{center}
\end{figure}

The brane configuration for the orbifold $\C^3/\Delta(3n^2)$
is obtained from that for $\C^3/\ZnZn$ by a $\Z_3$ quotient.
The $\Z_3$ acts as a $2\pi/3$ rotation
along the line $x^2=x^4=x^6$.
The brane configuration is given in Figure \ref{fig:brane27-3}.

The reason that the configuration gives the structure of the
irreducible representations of $\Delta(3n^2)$ is the same
as in \cite{Muto3}.
The gauge group coming from each box depend on
whether the box include fixed points of the $\Z_3$ action.
When $n/3$ is not an integer,
only the boxes with index (0,0) include fixed line of the $\Z_3$ action.
When $n/3$ is an integer, the boxes with indices
$(2n/3,n/3)$ and $(n/3,2n/3)$ also include fixed lines in addition to (0,0).
For such boxes, we must take a $\Z_3$ quotient,
which leads the gauge group to be $U(1)^3$
since the gauge field on the box takes the following form,
\begin{equation}
A \sim \left(
\begin{array}{ccc}
a_1&a_2&a_3 \\
a_3&a_1&a_2 \\
a_2&a_3&a_1
\end{array}
\right).
\end{equation}
It implies that such boxes correspond to a sum of three
one-dimensional representations of $\Delta(3n^2)$.
On the other hand,
for the boxs which do not include the fixed lines of the
$\Z_3$ action, three D-branes simply pile up
and give the gauge group $U(3)$.
It implies that such boxes correspond to three-dimensional
irreducible representations of $\Delta(3n^2)$.
One can see that the quotienting procedure precicely reproduces
the structure of the irreducible representations of $\Delta(3n^2)$.
We can also verify that matter contents obtained after the $\Z_3$ quotient
coincide with those specified by the quiver diagram of $\Delta(3n^2)$.

\subsection{Brane cube configurations for $\C^3/\Delta(6n^2)$}

\begin{figure}[htdp]
\begin{center}
\leavevmode
\epsfysize=40mm
\epsfbox{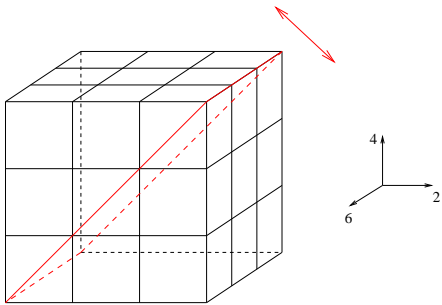}
\caption{The brane cube configuration for $\C^3/\Delta(6n^2)$ with $n=3$.
It is obtained from the brane configuration for $\C^3/\Delta(3n^2)$
by a $\Z_2$ quotient. The $\Z_2$ acts as a reflection with respect to the
plane $x_2=x_4$.}
\label{fig:brane54-3}
\end{center}
\end{figure}

The brane configuration for the orbifold $\C^3/\Delta(6n^2)$
is obtained from that for $\C^3/\Delta(3n^2)$ by a $\Z_2$
quotient.
The $\Z_2$ acts as a reflection with respect to the plane $x_2=x_4$
as indicated in the quiver diagram of $\Delta(6n^2)$
in Figure \ref{fig:quiver96}(b).
Combining with the $\Z_3$ identification for the configuration
corresponding to $\C^3/\Delta(3n^2)$,
it is equivalent to the reflections with respect to the planes $x_4=x_6$
and $x_6=x_2$.
The brane configuration is given in Figure \ref{fig:brane54-3}.
One can verify that the configuration reproduces the structure
of the quiver diagrams of $\Delta(6n^2)$.
For details on the structure of the irreducible representations,
see \cite{Muto3}.

\section{Discussions}
\reseteqnum

In this paper, we have proposed brane cube configurations
corresponding to D1-branes on nonabelian orbifolds
${\bf C}^3/\Gamma$ with $\Gamma \in SU(3)$.
This is based on the three-dimensional version of the
quiver diagrams of $\Gamma$.
The configuration consists of three kinds of NS5-branes
and D4-branes.
Since the D4-branes are bounded by NS5-branes along three directions,
the low energy theory becomes two-dimensional.
Due to the fact that the configuration is uniform
along a diagonal direction,
supersymmetry of the field theory is enhanced
twice as many as naively expected.

The original motivation to consider configurations
with three-dimensional structure comes from the study
of configurations for an orbifold $\C^3/\Gamma$
with $\Gamma$ an E-type subgroup of $SU(3)$.
As mentioned in introduction,
the quiver diagram of $\Sigma(648)$ is a quotient
of the quiver diagram of $\Delta(3 \times 3^2)$
by the tetrahedral group.
According to the correspondence between quiver diagrams
and brane configurations for $\C^3/\Gamma$,
we expect that a brane configuration for $\C^3/\Sigma(648)$
is obtained from the configuration for $\C^3/\Delta(3 \times 3^2)$
given in Figure \ref{fig:brane27-3}
by taking a quotient by the tetrahedral group.
In fact, the configuration in Figure \ref{fig:brane27-3}
has structure of the tetrahedral group:
the configuration consists of cubes,
which are in a sence dual to tetrahedra,
so we can define an action of the tetrahedral
group on the configuration.
It is now under investigation
whether such a quotient actually reproduces
required properties of the gauge theories.

Finally we would like to compare the configurations
in this paper with those given in \cite{Muto3}.
The configuration given in \cite{Muto3}
consists of a web of $(p,q)$ 5-branes and D3-branes.
As the D3-branes are bounded along two-directions
by $(p,q)$ 5-branes, they provide two-dimensional field theory,
which coincides with the brane cube configurations.
Thus it seems that we should start with {\sl D1}-branes on
$\C^4/\Gamma \sim \C^3/\Gamma \times \C$
to realize $\Z_3$ symmetry of brane configurations.
From this viewpoint,
the fact that the brane box model gives four-dimensional theory
is owing to the speciality of the abelian case
in which an explicit $\Z_3$ symmetry is not necessary.

As we have discussed, the configurations
given in \cite{Muto3} have esentially two-dimensional structure,
while the brane cube configurations are three-dimensional.
It is interesting to see whether there is a relation like duality
between the two types of configurations.
If such a relation can be found,
it may be possible to construct brane configurations
corresponding to $\C^3/\Gamma$ with $\Gamma$ an E-type subgroup
based on the brane configurations
made out of a web of $(p,q)$ 5-branes.

In \cite{Muto3}, it was also argued that
the brane configurations are dual to toric diagrams of $\C^3/\Gamma$
and that three-dimensional McKay correspondence
\cite{McKay}-\cite{IN} may be understood as T-duality.
We hope that the three-dimensional version of the quiver diagrams and
the brane cube configurations provide some hints
on investigations along these lines.

\vskip 1cm
\centerline{\large\bf Acknowledgements}

I would like to thank T. Kitao for valuable discussions.
This work is supported in part by Japan Society for the
Promotion of Science(No. 10-3815).

\end{document}